\documentclass[12pt]{iopart}

\usepackage{iopams}
\usepackage{setstack}
\usepackage{graphicx}
\usepackage{bm}
\usepackage{hyperref}
\hypersetup{colorlinks=false}

\begin{document}

\title[The Klein-Gordon equation in the spacetime of a charged and rotating black hole]{The Klein-Gordon equation in the spacetime of a charged and rotating black hole}

\author{V. B. Bezerra$^{1}$, H. S. Vieira$^{1}$ and Andr\'{e} A. Costa$^{2}$}

\address{$^{1}$ Departamento de F\'{i}sica, Universidade Federal da Para\'{i}ba, Caixa Postal 5008, CEP 58051-970, Jo\~{a}o Pessoa, PB, Brazil}
\address{$^{2}$ Instituto de F\'{i}sica, Universidade de S\~{a}o Paulo, Caixa Postal 66318, CEP 05315-970, S\~{a}o Paulo, SP, Brazil}

\ead{valdir@fisica.ufpb.br, horacio.santana.vieira@hotmail.com and alencar@if.usp.br}

\begin{abstract}
This work deals with the influence of the gravitational field produced by a charged and rotating black hole (Kerr-Newman spacetime) on massive scalar fields. We obtain an exact solution of the Klein-Gordon equation in this spacetime, which is given in terms of the confluent Heun functions. In the particular case corresponding to an extreme Kerr-Newman black hole the solution is given by the double confluent Heun functions. We also investigate the solutions in regions near the event horizon and far from the black hole.
\end{abstract}




\maketitle


\section{Introduction}
The study of the interaction of scalar and spinor particles with a gravitational fields goes back to the beginning of the last century, when the generalization of quantum mechanics to curved spaces was discussed motivated by the idea of constructing a theory combining quantum physics and general relativity. Along this line of research the study of solutions of the Klein-Gordon equation in some gravitational fields as well as their consequences have been discussed in the literature \cite{bezerra,pimentel,vakili,semiz,rowan,kraori,stephenson,rowan2,wu,elizalde,pimentel2,sandro}.

By the middle of 70's of last century, Rowan and Stephenson \cite{rowan}, solved the Klein-Gordon equation for a massive field in the Kerr-Newman spacetime. The obtained solutions are not valid for the whole spacetime, but only near the exterior horizon and at infinity, and are given in terms of Whittaker functions.

Solutions to the Klein-Gordon equation for a charged massive scalar field in the Kerr-Newman spacetime were obtained by Wu and Cai \cite{wu}. They showed that the radial and angular equations correspond to a generalized spin-weighted spheroidal wave functions, in the non-extreme case. In this context, the general solutions in integral forms and in power series expansion were obtained and the appropriate forms of the equations suitable to study some problems of physical interest, as the ones concerning black hole evaporation, black hole radiation and scattering states, among others, were presented and their solutions given. In the extreme case, the solutions of the radial equation were written in power series expansion and some discussion presented.

Another paper by Furuhashi and Nambu \cite{furuhashi} presents solutions of the Klein-Gordon equation for a massive scalar field in Kerr-Newman spacetime. The solutions were considered very far from the exterior event horizon and near the black hole event horizon. In the first case, it was found that the solutions regular at infinity is given in terms of the confluent hypergeometric functions, and in the opposite situation, namelly, near the horizon, the solution is given in terms of the Gauss hypergeometric function. Therefore, the obtained solutions are valid in a restricted ranges.

In our paper we obtain the exact solutions of the Klein-Gordon equation in the background under consideration, valid in the whole space which corresponds to the black hole exterior which means between the event horizon and infinity. In this sense, we extend the range in which the solutions are valid as compared with the ones obtained by Rowan and Stephenson \cite{rowan}. They are given in terms of solutions of the Heun equations \cite{ronveaux}. We also analyze the asymptotic behavior of the solutions and compared with the corresponding ones obtained by Rowan and Stephenson \cite{rowan}. The solution of the Klein-Gordon equation for the extreme Kerr-Newman black hole is also obtained and is given by the double confluent Heun function.

This paper is organized as follows. In the section II we introduce the Klein-Gordon equation in a curved background and write it in the Kerr-Newman spacetime, by separating the angular and radial parts. In section III we present the exact solution of the radial equation and discuss the asymptotic limits. In section IV, we consider the extreme Kerr-Newman spacetime and present the solution of the Klein-Gordon equation. Finally, in section V we present our conclusions.

\section{The Klein-Gordon equation in a Kerr-Newman spacetime}
We want to study the behavior of scalar fields, in the gravitational field of a charged and rotating black hole. In this way, we must solve the covariant Klein-Gordon equation, which is the equation that describes the behavior of scalar particles, in the curved spacetime under consideration which in our case is the one generated by a charged and rotating black hole. In a curved spacetime, we can write the Klein-Gordon equation in the covariant form which is given by
\begin{equation}
\left[\frac{1}{\sqrt{-g}}\partial_{\sigma}\left(g^{\sigma\tau}\sqrt{-g}\partial_{\tau}\right)+\mu^{2}\right]\Psi=0\ ,
\label{eq:Klein-Gordon_cova}
\end{equation}
On the other hand, the metric generated by a charged and rotating black hole is the Kerr-Newman metric \cite{misner}, which in the Boyer-Lindquist coordinates \cite{boyer} can be written as
\begin{eqnarray}
ds^{2} & = & \frac{\Delta}{\rho^{2}}\left(dt-a\sin^{2}\theta\ d\phi\right)^{2}-\frac{\sin^2\theta}{\rho^2}\left[\left(r^2+a^2\right)d\phi-a\ dt\right]^{2}\nonumber\\
& - & \frac{\rho^2}{\Delta}dr^{2}-\rho^{2}\ d\theta^{2}
\label{eq:metrica_Kerr-Newman}
\end{eqnarray}
where
\numparts
\begin{equation}
\Delta=r^{2}-2Mr+a^{2}+Q^{2}\ ,
\end{equation}
\begin{equation}
\rho^{2}=r^{2}+a^{2}\cos^{2}\theta\ .
\end{equation}
\endnumparts
Thus, writing down the Klein-Gordon equation in the Kerr-Newman spacetime given by (\ref{eq:metrica_Kerr-Newman}), we obtain the following equation
\begin{eqnarray}
& & \left\{\frac{1}{\Delta}\left[\left(r^{2}+a^{2}\right)^{2}-\Delta a^{2}\sin^{2}\theta\right]\frac{\partial^{2}}{\partial t^{2}}-\frac{\partial}{\partial r}\left(\Delta\frac{\partial}{\partial r}\right)\right.\nonumber\\
& - & \frac{1}{\sin\theta}\frac{\partial}{\partial\theta}\left(\sin\theta\frac{\partial}{\partial\theta}\right)-\frac{1}{\Delta\sin^{2}\theta}\left(\Delta-a^{2}\sin^{2}\theta\right)\frac{\partial^{2}}{\partial\phi^{2}}\nonumber\\
& - & \left.\frac{2a}{\Delta}\left[\Delta-\left(r^{2}+a^{2}\right)\right]\frac{\partial^{2}}{\partial\phi\partial t}+\rho^{2}\mu^{2}\right\}\Psi=0\ .
\label{eq:mov}
\end{eqnarray}

To solve the Eq.~(\ref{eq:mov}), we assume that its solution can be separated as follows
\begin{equation}
\Psi=\Psi(\mathbf{r},t)=R(r)S(\theta)\mbox{e}^{im\phi}\mbox{e}^{-i\omega t}\ .
\label{eq:separacao_variaveis}
\end{equation}
Substituting Eq.~(\ref{eq:separacao_variaveis}) into (\ref{eq:mov}), we find that
\begin{eqnarray}
& & \frac{1}{R}\frac{d}{dr}\left(\Delta\frac{dR}{dr}\right)+\frac{1}{S}\frac{1}{\sin\theta}\frac{d}{d\theta}\left(\sin\theta\frac{dS}{d\theta}\right)+\frac{\left(\Delta-a^{2}\sin^{2}\theta\right)\left(-m^{2}\right)}{\Delta\sin^{2}\theta}\nonumber\\
& + & \frac{2a\left[\Delta-\left(r^{2}+a^{2}\right)\right]m\omega}{\Delta}-\frac{1}{\Delta}\left[\left(r^{2}+a^{2}\right)^{2}-\Delta a^{2}\sin^{2}\theta\right]\left(-\omega^{2}\right)\nonumber\\
& - & \rho^{2}\mu^{2}=0\ .
\label{eq:mov_separavel}
\end{eqnarray}
This equation can be separated according to
\begin{equation}
\frac{1}{\sin\theta}\frac{d}{d\theta}\left(\sin\theta\frac{dS}{d\theta}\right)+\left(\lambda_{lm}+c^{2}\cos^{2}\theta-\frac{m^{2}}{\sin^{2}\theta}\right)S=0
\label{eq:mov_angular}
\end{equation}
where $c^{2}=a^{2}(\omega^{2}-\mu^{2})$, and
\begin{eqnarray}
& & \Delta\frac{d}{dr}\left(\Delta\frac{dR}{dr}\right)+[\omega^{2}\left(r^{2}+a^{2}\right)^{2}-4Ma\omega mr+2Q^{2}a\omega m-\mu^{2}r^{2}\Delta\nonumber\\
& + & m^{2}a^{2}-\left(\omega^{2}a^{2}+\lambda_{lm}\right)\Delta]R=0\ .
\label{eq:mov_radial_1}
\end{eqnarray}
The Eq.~(\ref{eq:mov_angular}) has as its solutions the oblate spheroidal harmonic functions $S_{lm}(ic,\cos\theta)$ with eigenvalues $\lambda_{lm}$, where $l,m$ are integers such that $|m|\leq l$ \cite{ford,morse}. What about the solution of the radial equation given by (\ref{eq:mov_radial_1})? As we can see, the radial equation has a complicated dependence with the coordinate $r$, which makes the determination of the solution not so simple.

In order to solve the radial part of the Klein-Gordon equation, let us write
\begin{equation}
\Delta=r^{2}-2Mr+a^{2}+Q^{2}=(r-r_{+})(r-r_{-})
\label{eq:Delta_Kerr-Newman}
\end{equation}
where
\begin{equation}
r_{\pm}=M\pm\left[M^{2}-\left(a^{2}+Q^{2}\right)\right]^{1/2}\ ,
\label{eq:raios}
\end{equation}
are the roots of $\Delta$ and correspond to the event and Cauchy horizons of the black hole. Using (\ref{eq:raios}), we can rewrite Eq.~(\ref{eq:mov_radial_1}) as
\begin{eqnarray}
& & (r-r_{+})(r-r_{-})\frac{d}{dr}\left[(r-r_{+})(r-r_{-})\frac{dR}{dr}\right]+[\omega^{2}\left(r^{2}+a^{2}\right)^{2}\nonumber\\
& - & 4Ma\omega mr+2Q^{2}a\omega m-\mu^{2}r^{2}(r-r_{+})(r-r_{-})+m^{2}a^{2}\nonumber\\
& - & \left(\omega^{2}a^{2}+\lambda_{lm}\right)(r-r_{+})(r-r_{-})]R=0\ .
\label{eq:mov_radial_2}
\end{eqnarray}
Now, we define a new coordinate $x$ and a parameter $d$ given by \cite{rowan}
\numparts
\begin{equation}
Mx=r-r_{+}\ ,
\label{eq:x_Kerr-Newman}
\end{equation}
\begin{equation}
2Md=r_{+}-r_{-}=2[M^{2}-(a^{2}+Q^{2})]^{1/2}\ .
\label{eq:d_Kerr-Newman}
\end{equation}
\endnumparts
The new coordinate is such that when $r \rightarrow r_{+}$, $x \rightarrow 0$, and when $r \rightarrow \infty$, we obtain that $x \rightarrow \infty$. We are interested to study the behavior of the particles only in the exterior region of the event horizon, namely, for $r > r_{+}$, which correspond to $0 \leq x < \infty$, in terms of the new coordinate. Thus, writing Eq.~(\ref{eq:mov_radial_2}) using the new coordinate, we get
\begin{eqnarray}
& & \frac{d}{dx}\left[x(x+2d)\frac{dR}{dx}\right]+\lshad\omega^{2}\frac{\{M^{2}[(x+d+1)^{2}-(d^{2}-1)]-Q^{2}\}^{2}}{M^{2}x(x+2d)}\nonumber\\
& - & \frac{4a\omega m(x+d+1)}{x(x+2d)}+\frac{2Q^{2}a\omega m}{M^{2}x(x+2d)}-M^{2}\mu^{2}(x+d+1)^{2}\nonumber\\
& + & \frac{m^{2}a^{2}}{M^{2}x(x+2d)}-(\omega^{2}a^{2}+\lambda_{lm})\rshad R=0\ .
\label{eq:mov_radial_3}
\end{eqnarray}

Defining a new function by $R(x)=Z(x)[x(x+2d)]^{-1/2}$, we can write Eq.~(\ref{eq:mov_radial_3}) in the following form
\begin{eqnarray}
& & \frac{d^{2}Z}{dx^{2}}+\left(M^{2}\right.\left(\omega^{2}-\mu^{2}\right)+\frac{1}{M^{2}x^{2}(x+2d)^{2}}\lshad\omega^{2}\left\{M^{4}\left[4(x+d+1)^{2}\right.\right.\nonumber\\
& + & \left.\left.4(x+d+1)x(x+2d)\right]-2M^{2}Q^{2}[x(x+2d)+2(x+d+1)]+Q^{4}\right\}\nonumber\\
& - & 4a\omega mM^{2}(x+d+1)+2Q^{2}a\omega m-\mu^{2}M^{4}\left[2x+(d+1)^{2}\right]x(x+2d)\nonumber\\
& + & \left.m^{2}a^{2}-\left(\omega^{2}a^{2}+\lambda_{lm}\right)M^{2}(x+2d)x+M^{2}d^{2}\rshad\right)Z=0\ .
\label{eq:mov_radial_4}
\end{eqnarray}
This differential equation can be written in a form which permits us to compare with a Heun differential equation, and this will be done in the next section.

\section{Solution of the Klein-Gordon equation in a Kerr-Newman spacetime}
In order to obtain the radial solution of the Klein-Gordon equation in a Kerr-Newman spacetime, given by Eq.~(\ref{eq:mov_radial_4}), we follow the same procedure adopted by Rowan and Stephenson \cite{rowan} and write the radial equation in the partial fraction form as
\begin{eqnarray}
\frac{d^{2}Z}{dx^{2}}+\left[M^{2}(\omega^{2}-\mu^{2})+\frac{1}{M^{2}}\left(\frac{A}{x^{2}}+\frac{B}{x}+\frac{C}{(x+2d)^{2}}+\frac{D}{(x+2d)}\right)\right]Z=0\ ,\nonumber\\
\label{eq:radial_Kerr-Newman}
\end{eqnarray}
where the coefficients $A$, $B$, $C$ and $D$ are given by
\numparts
\begin{eqnarray}
A & = & \frac{1}{4d^{2}}\left[m^{2}a^{2}+2Q^{2}a\omega m-4a\omega mM^{2}(d+1)-4M^{2}\omega^{2}Q^{2}(d+1)\right.\nonumber\\
& + & \left.4M^{4}\omega^{2}(d+1)^{2}+M^{2}d^{2}+\omega^{2}Q^{4}\right]\ ,
\label{eq:A}
\end{eqnarray}
\begin{eqnarray}
B & = & \frac{1}{4d^{3}}\left[-m^{2}a^{2}-2Q^{2}a\omega m+4a\omega mM^{2}-4M^{2}\omega^{2}Q^{2}(d^{2}-1)\right.\nonumber\\
& + & 4M^{4}\omega^{2}(d+1)^{2}(2d-1)-2M^{4}\mu^{2}d^{2}(d+1)^{2}\nonumber\\
& - & \left.2M^{2}d^{2}(\omega^{2}a^{2}+\lambda_{lm})-M^{2}d^{2}-\omega^{2}Q^{4}\right]\ ,
\label{eq:B}
\end{eqnarray}
\begin{eqnarray}
C & = & \frac{1}{4d^{2}}\left[m^{2}a^{2}+2Q^{2}a\omega m+4a\omega mM^{2}(d-1)+4M^{2}\omega^{2}Q^{2}(d-1)\right.\nonumber\\
& + & \left.4M^{4}\omega^{2}(d-1)^{2}+M^{2}d^{2}+\omega^{2}Q^{4}\right]\ ,
\label{eq:C}
\end{eqnarray}
\begin{eqnarray}
D & = & \frac{1}{4d^{3}}\left[m^{2}a^{2}+2Q^{2}a\omega m-4a\omega mM^{2}+4M^{2}\omega^{2}Q^{2}(d^{2}-1)\right.\nonumber\\
& + & 4M^{4}\omega^{2}(d-1)^{2}(2d+1)+2M^{4}\mu^{2}d^{2}(d-1)^{2}\nonumber\\
& + & \left.2M^{2}d^{2}(\omega^{2}a^{2}+\lambda_{lm})+M^{2}d^{2}+\omega^{2}Q^{4}\right]\ .
\label{eq:D}
\end{eqnarray}
\endnumparts
Now, we define a new independent variable, $z$, such that
\begin{equation}
z=-\frac{1}{2}\frac{x}{d}\ .
\label{eq:variavel_z_Kerr-Newman}
\end{equation}
Changing once more the variable, from $x$ to $z$, Eq.~(\ref{eq:radial_Kerr-Newman}) can be written as
\begin{eqnarray}
& & \frac{d^{2}Z}{dz^{2}}+\left[4d^{2}M^{2}(\omega^{2}-\mu^{2})+\frac{A}{M^{2}z^{2}}+\frac{-2dB}{M^{2}z}+\frac{C}{M^{2}(z-1)^{2}}\right.\nonumber\\
& + & \left.\frac{-2dD}{M^{2}(z-1)}\right]Z=0\ .
\label{eq:radial_Kerr-Newman_normal}
\end{eqnarray}

Consider a linear ordinary differential equation of second order in the standard form
\begin{equation}
\frac{d^{2}U}{dz^{2}}+p(z)\frac{dU}{dz}+q(z)U=0\ .
\label{eq:EDO_forma_padrao}
\end{equation}
Changing the function $U(z)$ using the relation
\begin{equation}
U(z)=Z(z)\mbox{e}^{-\frac{1}{2}\int p(z)dz}\ ,
\label{eq:U}
\end{equation}
Eq.~(\ref{eq:EDO_forma_padrao}) turns into the normal form
\begin{equation}
\frac{d^2Z}{dz^2}+I(z)Z=0\ ,
\label{eq:EDO_forma_normal}
\end{equation}
where
\begin{equation}
I(z)=q(z)-\frac{1}{2}\frac{dp(z)}{dz}-\frac{1}{4}\left[p(z)\right]^2\ .
\label{eq:I}
\end{equation}
Now, let us consider the confluent Heun equation \cite{fiziev}
\begin{equation}
\frac{d^{2}U}{dz^{2}}+\left(\alpha+\frac{\beta+1}{z}+\frac{\gamma+1}{z-1}\right)\frac{dU}{dz}+\left(\frac{\mu}{z}+\frac{\nu}{z-1}\right)U=0\ ,
\label{eq:Heun_confluente_forma_canonica}
\end{equation}
where $U(z)=\mbox{HeunC}(\alpha,\beta,\gamma,\delta,\eta,z)$ are the confluent Heun functions, with the parameters $\alpha$, $\beta$, $\gamma$, $\delta$ and $\eta$, related to $\mu$ and $\nu$ by
\begin{equation}
\mu=\frac{1}{2}(\alpha-\beta-\gamma+\alpha\beta-\beta\gamma)-\eta\ ,
\label{eq:mu_Heun_conlfuente_2}
\end{equation}
\begin{equation}
\nu=\frac{1}{2}(\alpha+\beta+\gamma+\alpha\gamma+\beta\gamma)+\delta+\eta\ ,
\label{eq:nu_Heun_conlfuente_2}
\end{equation}
according to the standard package of the \textbf{Maple}\texttrademark \textbf{17}. Using Eqs.~(\ref{eq:EDO_forma_padrao})-(\ref{eq:I}), we can write Eq.~(\ref{eq:Heun_confluente_forma_canonica}) in the normal form as \cite{ronveaux}
\begin{equation}
\frac{d^{2}Z}{dz^{2}}+\left[B_{1}+\frac{B_{2}}{z^{2}}+\frac{B_{3}}{z}+\frac{B_{4}}{(z-1)^{2}}+\frac{B_{5}}{z-1}\right]Z=0
\label{eq:Heun_confluente_forma_normal}
\end{equation}
where the coefficients $B_{1}$, $B_{2}$, $B_{3}$, $B_{4}$ and $B_{5}$ are given by
\numparts
\label{eq:B1-B5_Heun_confluente_forma_normal}
\begin{equation}
B_{1} \equiv -\frac{1}{4}\alpha^{2}\ ,
\end{equation}
\begin{equation}
B_{2} \equiv \frac{1}{4}(1-\beta^{2})\ ,
\end{equation}
\begin{equation}
B_{3} \equiv\frac{1}{2}(1-2\eta)\ ,
\end{equation}
\begin{equation}
B_{4} \equiv \frac{1}{4}(1-\gamma^{2})\ ,
\end{equation}
\begin{equation}
B_{5} \equiv \frac{1}{2}(-1+2\delta+2\eta)\ .
\end{equation}
\endnumparts

The radial part of the Klein-Gordon equation in the Kerr-Newman spacetime in the exterior region of the event horizon, given by (\ref{eq:radial_Kerr-Newman_normal}), can be written as (\ref{eq:Heun_confluente_forma_normal}), and therefore, its solution is given by
\begin{equation}
Z(z)=U(z)\mbox{e}^{\frac{1}{2}\int\left(\alpha+\frac{\beta+1}{z}+\frac{\gamma+1}{z-1}\right)dz}\ ,
\label{eq:solZ}
\end{equation}
where $U(z)$ is solution of the confluent Heun equation (\ref{eq:Heun_confluente_forma_canonica}), and the parameters $\alpha$, $\beta$, $\gamma$, $\delta$ and $\eta$ are obtained from the following relations
\numparts
\label{eq:B1-B5_radial_Kerr-Newman_normal}
\begin{equation}
-\frac{1}{4}\alpha^{2}=4d^{2}M^{2}(\omega^{2}-\mu^{2})\ ,
\end{equation}
\begin{equation}
\frac{1}{4}(1-\beta^{2})=\frac{A}{M^{2}}\ ,
\end{equation}
\begin{equation}
\frac{1}{2}(1-2\eta)=\frac{-2dB}{M^{2}}\ ,
\end{equation}
\begin{equation}
\frac{1}{4}(1-\gamma^{2})=\frac{C}{M^{2}}\ ,
\end{equation}
\begin{equation}
\frac{1}{2}(-1+2\delta+2\eta)=\frac{-2dD}{M^{2}}\ .
\end{equation}
\endnumparts
Thus, from the above relations, we find that
\numparts
\begin{equation}
\alpha=4dM(\mu^{2}-\omega^{2})^{1/2}\ ,
\label{eq:alpha_HeunC_Kerr-Newman}
\end{equation}
\begin{equation}
\beta=\sqrt{1-\frac{4A}{M^{2}}}\ ,
\label{eq:beta_HeunC_Kerr-Newman}
\end{equation}
\begin{equation}
\gamma=\sqrt{1-\frac{4C}{M^{2}}}\ ,
\label{eq:gamma_HeunC_Kerr-Newman}
\end{equation}
\begin{equation}
\delta=-\frac{2d}{M^{2}}(B+D)\ ,
\label{eq:delta_HeunC_Kerr-Newman}
\end{equation}
\begin{equation}
\eta=\frac{1}{2}+\frac{2dB}{M^{2}}\ .
\label{eq:eta_HeunC_Kerr-Newman}
\end{equation}
\endnumparts
The general solution of Eq.~(\ref{eq:mov_radial_2}) over the entire range $0 \leq z < \infty$ is obtained with the use of Eq.~(\ref{eq:solZ}). It is given by
\begin{eqnarray}
R(z) & = & \frac{M}{\Delta^{1/2}}\mbox{e}^{\frac{1}{2}\alpha z}(z-1)^{\frac{1}{2}(1+\gamma)}z^{\frac{1}{2}(1+\beta)}\nonumber\\
& \times & \{c_{1}\ \mbox{HeunC}(\alpha,\beta,\gamma,\delta,\eta,z)+c_{2}\ z^{-\beta}\ \mbox{HeunC}(\alpha,-\beta,\gamma,\delta,\eta,z)\}\ ,\nonumber\\
\label{eq:solucao_geral_radial_Kerr-Newman}
\end{eqnarray}
where $\mbox{HeunC}(\alpha,\pm\beta,\gamma,\delta,\eta,z)$ are the confluent Heun functions, $c_{1}$ and $c_{2}$ are constants and $\alpha$, $\beta$, $\gamma$, $\delta$ and $\eta$ are fixed by relations (\ref{eq:alpha_HeunC_Kerr-Newman})-(\ref{eq:eta_HeunC_Kerr-Newman}). These two functions form linearly independent solutions of the confluent Heun differential equation provided $\beta$ is not integer. However, there is no any specific physical reason to impose that $\beta$ should be integer.

Now, let us analyze the asymptotic behavior of the general solution of Eq.(\ref{eq:mov_radial_2}), given by Eq.(\ref{eq:solucao_geral_radial_Kerr-Newman}), over the range $0 \leq x < \infty$. Firstly, we will consider a region close to the event horizon, which means that $r \rightarrow r_{+}$, that is, $x \rightarrow 0$. Secondly, we will consider a region very far from the black hole, that is, $r \rightarrow \infty$, or equivalently, $x \rightarrow \infty$.

To take these limits into consideration is important in order to obtain the appropriate solutions to study some aspects of black hole radiation, in which case we need to know the outgoing wave in the horizon surface $r=r_{+}$.

\subsection{Case 1. \texorpdfstring{$x \rightarrow 0$}{x->0}}
When $z \rightarrow 0$ we have that $x \rightarrow 0$ and $r \rightarrow r_{+}$. Thus, using the expansion in power series for all $z$ of the confluent Heun function, namely
\begin{eqnarray}
\mbox{HeunC}(\alpha,\beta,\gamma,\delta,\eta,z) & = & 1+\frac{1}{2}\frac{(-\alpha\beta+\beta\gamma+2\eta-\alpha+\beta+\gamma)}{(\beta+1)}z\nonumber\\
& + & \frac{1}{8}\frac{1}{(\beta+1)(\beta+2)}\left(\alpha^{2}\beta^{2}\right.-2\alpha\beta^{2}\gamma+\beta^{2}\gamma^{2}\nonumber\\
& - & 4\eta\alpha\beta+4\eta\beta\gamma+4\alpha^{2}\beta-2\alpha\beta^{2}-6\alpha\beta\gamma\nonumber\\
& + & 4\beta^{2}\gamma+4\beta\gamma^{2}+4\eta^{2}-8\eta\alpha+8\eta\beta+8\eta\gamma\nonumber\\
& + & 3\alpha^{2}-4\alpha\beta-4\alpha\gamma+3\beta^{2}+4\beta\delta\nonumber\\
& + & \left.10\beta\gamma+3\gamma^{2}+8\eta+4\beta+4\delta+4\gamma\right)z^2+...\ ,
\label{eq:serie_HeunC_todo_z}
\end{eqnarray}
the solutions of (\ref{eq:mov_radial_2}), in that limit, are given by
\begin{equation}
R(r)\sim\left\{
\begin{array}{l}
	\frac{M}{\Delta^{1/2}}\left[-\frac{1}{2dM}(r-r_{+})\right]^{\frac{1}{2}+\bar{m}}\ ,\\
	\\
	\frac{M}{\Delta^{1/2}}\left[-\frac{1}{2dM}(r-r_{+})\right]^{\frac{1}{2}-\bar{m}}\ ,
\end{array}
\right.
\label{eq:solucao_radial_caso1_limite}
\end{equation}
with $\bar{m}$ defined by
\begin{equation}
\bar{m}^{2}=\frac{1}{4}-\frac{A}{M^{2}}\ .
\label{eq:m_caso1}
\end{equation}

At this point, we can compare the obtained result with the ones of Rowan and Stepheson \cite{rowan}. It is worth calling attention to the difference between the functional form of the asymptotic behavior of the general solution of the radial equation obtained analytically, given by Eq.(\ref{eq:solucao_radial_caso1_limite}), and the approximated solution obtained by Rowan and Stephenson \cite{rowan}, given by
\begin{equation}
R(r)\sim\left\{
\begin{array}{l}
	\frac{M}{\Delta^{1/2}}\mbox{e}^{-(\sqrt{F}/M)(r-r_{+})}\left[\frac{2\sqrt{F}}{M}(r-r_{+})\right]^{\frac{1}{2}+\bar{m}}\ ,\\
	\\
	\frac{M}{\Delta^{1/2}}\mbox{e}^{-(\sqrt{F}/M)(r-r_{+})}\left[\frac{2\sqrt{F}}{M}(r-r_{+})\right]^{\frac{1}{2}-\bar{m}}\ ,
\end{array}
\right.
\label{eq:solucao_radial_caso1}
\end{equation}
where
\begin{equation}
F=M^{2}(\mu^{2}-\omega^{2})-\frac{C}{4M^{2}d^{2}}-\frac{D}{2M^{2}d}\ .
\label{eq:F}
\end{equation}
It is worth noticing that expanding (\ref{eq:solucao_radial_caso1}) up to first order in terms of $x=(r-r_{+})/M$, we conclude that the results agree with ours, given by Eq.~(\ref{eq:solucao_radial_caso1_limite}), except for a multiplicative constant, which should be adjusted appropriately.

\subsection{Case 2. \texorpdfstring{$x \rightarrow \infty$}{x->+infinity}}
When $|z| \rightarrow \infty$ we have that $|x| \rightarrow \infty$ and $r \rightarrow \infty$. Thus, using the fact that in the neighborhood of the irregular singular point at infinity, the two solutions of the confluent Heun equation exist, in general they can be expanded (in a sector) in the following asymptotic series \cite{ronveaux}
\begin{equation}
\mbox{HeunC}(\alpha,\beta,\gamma,\delta,\eta,z) \sim \left\{
\begin{array}{l}
	z^{-\left(\frac{\beta+\gamma+2}{2}+\frac{\delta}{\alpha}\right)}\ ,\\
	\\
	z^{-\left(\frac{\beta+\gamma+2}{2}-\frac{\delta}{\alpha}\right)}\mbox{e}^{-\alpha z}\ ,
\end{array}
\right.
\label{eq:assintotica_HeunC_z_grande}
\end{equation}
where we are keeping only the first term of this power-series asymptotics. Thus, the solutions of (\ref{eq:mov_radial_2}) in this limit are given by
\begin{equation}
R(r)\sim\left\{
\begin{array}{l}
	\frac{M}{\Delta^{1/2}}\mbox{e}^{-\left[(\mu^{2}-\omega^{2})^{1/2}(r-r_{+})\right]}\left[-\frac{1}{2dM}(r-r_{+})\right]^{\kappa}\ ,\\
	\\
	\frac{M}{\Delta^{1/2}}\mbox{e}^{+\left[(\mu^{2}-\omega^{2})^{1/2}(r-r_{+})\right]}\left[-\frac{1}{2dM}(r-r_{+})\right]^{-\kappa}\ ,
\end{array}
\right.
\label{eq:solucao_radial_caso2_limite}
\end{equation}
with $\kappa$ defined by
\begin{equation}
\kappa=\frac{B+D}{2M^{3}(\mu^{2}-\omega^{2})^{1/2}}\ .
\label{eq:kappa_caso2}
\end{equation}

Therefore, at infinity, we have asymptotic forms which are consistent with the fact that very far from the black hole, the Kerr-Newman spacetime tends to Minkowski spacetime.

In this case there is also a difference between Eq.(\ref{eq:solucao_radial_caso2_limite}) and the one obtained by Rowan and Stephenson \cite{rowan}, given by
\begin{equation}
R(r)\sim\left\{
\begin{array}{l}
	\frac{M}{\Delta^{1/2}}\mbox{e}^{-\left[(\mu^{2}-\omega^{2})^{1/2}(r-r_{+})\right]}\left[2\left(\mu^{2}-\omega^{2}\right)^{1/2}(r-r_{+})\right]^{\kappa}\ ,\\
	\\
	\frac{M}{\Delta^{1/2}}\mbox{e}^{+\left[(\mu^{2}-\omega^{2})^{1/2}(r-r_{+})\right]}\left[-2\left(\mu^{2}-\omega^{2}\right)^{1/2}(r-r_{+})\right]^{-\kappa}\ ,
\end{array}
\right.
\label{eq:solucao_radial_caso2}
\end{equation}
provided $\mu^{2}-\omega^{2} \neq 0$. Once more, the two results are equivalent, except for a multiplicative constant.

\section{Solution of the Klein-Gordon equation in an extreme Kerr-Newman spacetime}
Let us now analyze the special case (extreme case) in which $a^{2}=M^{2}-Q^{2}$, so that the metric given by Eq.(\ref{eq:metrica_Kerr-Newman}) reduces to the form of the extreme Kerr-Newman spacetime. We should notice that the radial part of the Klein-Gordon equation, given by Eq.~(\ref{eq:radial_Kerr-Newman}), does not admits this case, because $a^{2}=M^{2}-Q^{2}$ implies, due to Eq.~(\ref{eq:d_Kerr-Newman}), that $d=0$ and, therefore, the coefficients $A$, $B$, $C$ and $D$ given by Eq.~(\ref{eq:A})-(\ref{eq:D}) are divergent. Thus, to study the extreme Kerr-Newman spacetime, we start from Eq.~(\ref{eq:mov_radial_4}) and, taking $a^{2}=M^{2}-Q^{2}$ and $d=0$, we obtain
\begin{eqnarray}
& & \frac{d^{2}Z}{dx^{2}}+\left(M^{2}\right.\left(\omega^{2}-\mu^{2}\right)+\frac{1}{M^{2}x^{4}}\lshad-4\omega mM^{3}(x+1)\left(M^{2}-Q^{2}\right)^{1/2}\nonumber\\
& + & m^{2}\left(M^{2}-Q^{2}\right)+\omega^{2}\left\{Q^{4}-2M^{2}Q^{2}\left[x^{2}+2(x+1)\right]\right.\nonumber\\
& + & \left.M^{4}\left[4(x+1)^{2}+4(x+1)x^{2}\right]\right\}-\mu^{2}M^{4}(2x+1)x^{2}\nonumber\\
& + & 2\omega mQ^{2}\left(M^{2}-Q^{2}\right)^{1/2}-\left[\omega^{2}\left(M^{2}-Q^{2}\right)+\lambda_{lm}\right]\left.M^{2}x^{2}\rshad\right)Z=0\ .
\label{eq:mov_radial_4_Kerr-Newman_extremo}
\end{eqnarray}
Now, we write Eq.(\ref{eq:mov_radial_4_Kerr-Newman_extremo}) in the partial fraction form as
\begin{equation}
\frac{d^{2}Z}{dx^{2}}+\left[F+\frac{A}{x}+\frac{B}{x^{2}}+\frac{C}{x^{3}}+\frac{D}{x^{4}}\right]Z=0\ ,
\label{eq:radial_Kerr-Newman_extremo}
\end{equation}
where the coefficients of the partial fractions are given by
\numparts
\label{A-D_Kerr-Newman_extremo}
\begin{equation}
A=4M^{2}\omega^{2}-2M^{2}\mu^{2}\ ,
\end{equation}
\begin{equation}
B=7M^{2}\omega^{2}-M^{2}\mu^{2}-\lambda_{lm}-Q^{2}\omega^{2}\ ,
\end{equation}
\begin{equation}
C=-4\omega\left[m\left(M^{2}-Q^{2}\right)^{1/2}-2M^{2}\omega+Q^{2}\omega\right]\ ,
\end{equation}
\begin{eqnarray}
D & = & \frac{m^{2}M^{2}-4\omega mM^{2}\left(M^{2}-Q^{2}\right)^{1/2}+4M^{4}\omega^{2}-m^{2}Q^{2}}{M^{2}}\nonumber\\
& + & \frac{2\omega mQ^{2}\left(M^{2}-Q^{2}\right)^{1/2}-4M^{2}Q^{2}\omega^{2}+Q^{4}\omega^{2}}{M^{2}}\ ,
\end{eqnarray}
\begin{equation}
F=M^{2}\left(\omega^{2}-\mu^{2}\right)\ .
\end{equation}
\endnumparts

Let us define a new independent variable, $z$, such that
\begin{equation}
z=\frac{i\left(FD^{3}\right)^{\frac{1}{4}}x-D}{i\left(FD^{3}\right)^{\frac{1}{4}}x+D}\ .
\label{eq:variavel_z_Kerr-Newman_extremo}
\end{equation}
Defining a new function given by $Z(z)=U(z)x^{1/2}$, where $x=x(z)$ is obtained from Eq.~(\ref{eq:variavel_z_Kerr-Newman_extremo}), we can rewrite Eq.~(\ref{eq:radial_Kerr-Newman_extremo}) in the normal form as
\begin{eqnarray}
& & \frac{d^{2}U}{dz^{2}}+\left\{\frac{\left[-B-\frac{iC\left(FD^{3}\right)^{\frac{1}{4}}}{2D}+\frac{iA\left(FD^{3}\right)^{\frac{3}{4}}}{2FD^{2}}\right]}{(z-1)}\right.\nonumber\\
& + & \frac{\left[B+\frac{iC\left(FD^{3}\right)^{\frac{1}{4}}}{2D}-\frac{iA\left(FD^{3}\right)^{\frac{3}{4}}}{2FD^{2}}\right]}{(z+1)}+\frac{\left[B-\frac{iA\left(FD^{3}\right)^{\frac{3}{4}}}{FD^{2}}\right]}{(z-1)^{2}}\nonumber\\
& + & \frac{\left[B+\frac{iC\left(FD^{3}\right)^{\frac{1}{4}}}{D}\right]}{(z+1)^{2}}+\frac{\left[\frac{2iA\left(FD^{3}\right)^{\frac{3}{4}}}{FD^2}\right]}{(z-1)^{3}}+\frac{\left[\frac{2iC\left(FD^{3}\right)^{\frac{1}{4}}}{D}\right]}{(z+1)^{3}}+\frac{\left[-\frac{4\left(FD^{3}\right)^{\frac{1}{2}}}{D}\right]}{(z-1)^{4}}\nonumber\\
& + & \left.\frac{\left[-\frac{4\left(FD^{3}\right)^{\frac{1}{2}}}{D}\right]}{(z+1)^{4}}\right\}U=0\ .
\label{eq:radial_Kerr-Newman_extremo_normal}
\end{eqnarray}

The double confluent Heun equation is given by
\begin{eqnarray}
& & \frac{d^{2}U}{dz^{2}}+\left[\frac{2z^{5}-\alpha z^{4}-4z^{3}+2z+\alpha}{(z-1)^{3}(z+1)^{3}}\right]\frac{dU}{dz}\nonumber\\
& + & \left[\frac{\beta z^{2}-(-\gamma-2\alpha)z+\delta}{(z-1)^{3}(z+1)^{3}}\right]U=0\ .
\label{eq:Heun_duplamente_confluente_forma_canonica}
\end{eqnarray}
Using Eqs.~(\ref{eq:EDO_forma_padrao})-(\ref{eq:I}) we can write Eq.~(\ref{eq:Heun_duplamente_confluente_forma_canonica}) in the normal form
\begin{eqnarray}
& & \frac{d^{2}Z}{dz^{2}}+\left[\frac{A}{(z-1)}+\frac{B}{(z+1)}+\frac{C}{(z-1)^{2}}+\frac{D}{(z+1)^{2}}\right.\nonumber\\
& + & \left.\frac{F}{(z-1)^{3}}+\frac{G}{(z+1)^{3}}+\frac{J}{(z-1)^{4}}+\frac{K}{(z+1)^{4}}\right]Z=0\ .
\label{eq:Heun_duplamente_confluente_forma_normal}
\end{eqnarray}
Thus, comparing Eq.~(\ref{eq:Heun_duplamente_confluente_forma_normal}) with the radial part of the Klein-Gordon equation in an extreme Kerr-Newman spacetime, given by (\ref{eq:radial_Kerr-Newman_extremo_normal}), we have that
\begin{equation}
Z(z)=U(z)\mbox{e}^{\frac{1}{2}\int\left[\frac{2z^{5}-\alpha z^{4}-4z^{3}+2z+\alpha}{(z-1)^{3}(z+1)^{3}}\right]dz}\ ,
\label{eq:solZextremo}
\end{equation}
where $U(z)$ is solution of the double confluent Heun differential equation given by (\ref{eq:Heun_duplamente_confluente_forma_canonica}). Now, the relations between the parameters are
\numparts
\begin{equation}
\frac{1}{32}(-8+\alpha^{2}-2\beta+6\delta)=-B-\frac{iC\left(FD^{3}\right)^{\frac{1}{4}}}{2D}+\frac{iA\left(FD^{3}\right)^{\frac{3}{4}}}{2FD^{2}}\ ,
\label{eq:A_radial_Kerr-Newman_extremo_normal}
\end{equation}
\begin{equation}
\frac{1}{32}(8-\alpha^{2}+2\beta-2\gamma-6\delta)=B-\frac{iA\left(FD^{3}\right)^{\frac{3}{4}}}{FD^{2}}\ ,
\label{eq:C_radial_Kerr-Newman_extremo_normal}
\end{equation}
\begin{equation}
\frac{1}{8}(\beta+\gamma+\delta)=\frac{2iA\left(FD^{3}\right)^{\frac{3}{4}}}{FD^2}\ ,
\label{eq:F_radial_Kerr-Newman_extremo_normal}
\end{equation}
\begin{equation}
-\frac{1}{16}(-\alpha^{2})=-\frac{4\left(FD^{3}\right)^{\frac{1}{2}}}{D}\ .
\label{eq:J_radial_Kerr-Newman_extremo_normal}
\end{equation}
\endnumparts
Using Eqs.~(\ref{eq:A_radial_Kerr-Newman_extremo_normal})-(\ref{eq:J_radial_Kerr-Newman_extremo_normal}) we can write, explicitly, the coefficients $\alpha$, $\beta$, $\gamma$ and $\delta$, as
\numparts
\label{eq:alpha-delta_HeunD_Kerr-Newman_extremo}
\begin{equation}
\alpha=\frac{8\left(FD^{3}\right)^{\frac{1}{4}}}{D^{1/2}}\ ,
\end{equation}
\begin{equation}
\beta=-1+4B-\frac{4iC\left(FD^{3}\right)^{\frac{1}{4}}}{D}+\frac{8\left(FD^{3}\right)^{\frac{1}{2}}}{D}+\frac{4iA\left(FD^{3}\right)^{\frac{3}{4}}}{FD^{2}}\ ,
\end{equation}
\begin{equation}
\gamma=\frac{8i\left[CFD\left(FD^{3}\right)^{\frac{1}{4}}+A\left(FD^{3}\right)^{\frac{3}{4}}\right]}{FD^{2}}\ ,
\end{equation}
\begin{equation}
\delta=1-4B-\frac{4iC\left(FD^{3}\right)^{\frac{1}{4}}}{D}+\frac{8\left(FD^{3}\right)^{\frac{1}{2}}}{D}+\frac{4iA\left(FD^{3}\right)^{\frac{3}{4}}}{FD^{2}}\ .
\end{equation}
\endnumparts

Therefore, taking into account (\ref{eq:solZextremo}), the general solution of Eq.~(\ref{eq:mov_radial_2}) over the range $0 \leq z < \infty$, for the case of the extreme Kerr-Newman metric, is given by
\begin{eqnarray}
R(z) & = & \frac{M}{\Delta^{1/2}}\mbox{e}^{\frac{1}{2}\frac{\alpha z}{(z-1)(z+1)}}\nonumber\\
& \times & \left\{c_{1}\ \mbox{HeunD}\left(-\alpha,-\delta,-\gamma,-\beta,\frac{1}{z}\right)\right.\nonumber\\
& + & \left.c_{2}\ \mbox{e}^{-\frac{\alpha z}{(z-1)(z+1)}}\ \mbox{HeunD}\left(\alpha,-\delta,-\gamma,-\beta,\frac{1}{z}\right)\right\}\ ,
\label{eq:solucao_geral_radial_Kerr-Newman_extremo}
\end{eqnarray}
where $c_{1}$ and $c_{2}$ are constants and the following identity for the double confluent Heun function was used:
\begin{equation}
\mbox{HeunD}\left(-\alpha,-\delta,-\gamma,-\beta,\frac{1}{z}\right)=\mbox{HeunD}(\alpha,\beta,\gamma,\delta,z)\ .
\label{eq:identidade_Heun_duplamente_confluente}
\end{equation}
Therefore, the solution for the extreme Kerr-Newman metric is different than the one obtained for the general Kerr-Newman metric.

\section{Conclusions}
In this paper, we presented analytic solutions of the radial part of the Klein-Gordon equation for a massive scalar field in the spacetime of a charged and rotating black hole (Kerr-Newman spacetime).

These solutions extend the ones obtained by Rowan and Stephenson in the sense that now we have analytic solutions for all spacetime, which means, in the region between the event horizon and infinity, contrary to the results of Rowan and Stephenson which were obtained for asymptotic regions, namely, very close to the horizons and far from the black hole. The solution is given in terms of the confluent Heun functions, and is valid over the range $0 \leq z < \infty$. In this way, the obtained solution generalizes the results found by Rowan and Stephenson \cite{rowan}.

Additionally, we also obtained the solutions to the extreme Kerr-Newman space\-time, which are given in terms of the double confluent Heun functions and are qualitativelly different from the one which solves the radial equation for the Kerr-Newman spacetime.

The obtained results have the advantage, as compared with the papers by Rowan and Stephenson \cite{rowan} and Furuhashi and Nambu \cite{furuhashi}, that the solutions are valid from the exterior event horizon to infinity, instead of to be valid only closed to the exterior event horizon and at infinity, as in Rowan and Stepheson \cite{rowan} and Furuhashi and Nambu \cite{furuhashi} papers. On the other hand, compared with the paper by Wu and Cai \cite{wu}, our results are given in terms of analytic functions, and not in integral forms or in form of series.

The obtained solutions certainly, will be important to study the physics of black hole radiation, scattering process, stationary state energy and in the analysis of stability \cite{furuhashi} of charged and uncharged black holes.

\ack The authors would like to thank Conselho Nacional de Desenvolvimento Cient\'{i}fico e Tecnol\'{o}gico (CNPq) for partial financial support.

\end{document}